\documentclass[11pt]{article}
\usepackage{moriond,epsfig,amsmath}

\def\Journal#1#2#3#4{{#1} {\bf #2}, #3 (#4)}

\def\AA{\em A\&A}

\def\be{\begin{equation}}
\def\ee{\end{equation}}
\def\bea{\begin{eqnarray}}
\def\eea{\end{eqnarray}}

\newcommand{\va}{\pmb{a}}
\newcommand{\vn}{\pmb{n}}
\newcommand{\vx}{\pmb{x}}

\newcommand{\tR}{{\bf R}}

\begin{document}
\vspace*{4cm}
\title{COMPONENT SEPARATION IN CMB OBSERVATIONS}

\author{ J. DELABROUILLE }

\address{CNRS -- Laboratoire APC, 10 rue A. Domon et L. Duquet \\
75013 Paris, France}

\maketitle\abstracts{
In these proceedings, we discuss the extraction, in WMAP 5 year data, of a clean CMB map, of foreground emission (dominated by emission of the interstellar medium of our galaxy), and of the tiny signal from Sunyaev Zel'dovich effect in the direction of known galaxy clusters. The implementation of an Internal Linear Combination method locally in both pixel and harmonic space, with the use of a decomposition of WMAP maps onto a frame of spherical needlets, permits to extract a full sky CMB map, with good accuracy even in regions close to the galactic plane. Proper subtraction of this estimated CMB from WMAP original observations provides us with CMB--free foreground maps, which can be used for the analysis of the emission of the galactic interstellar medium and for detecting and measuring emissions from compact sources.  Finally, while the Sunyaev-Zel'dovich cannot be detected for individual clusters in WMAP data, due to lack of spatial resolution and sensitivity, a stacking analysis of tentative detections towards a number of known ROSAT clusters permits to detect the SZ effect in WMAP data and measure how the SZ flux scales with cluster mass and X-ray luminosity.}

\section{Introduction}
The separation of emissions from different astrophysical or cosmological origin in Cosmic Microwave Background (CMB) observations
is a crucial aspect of CMB data analysis. Component separation consists in extracting maps (or any set of relevant parameters) describing the emission of individual components of interest, from a set of observations containing emissions from several such components \cite{review_jd_jfc}. For this purpose, one makes use of prior information about the morphology or the shape of individual components, about their color (emission law as a function of wavelength), about their statistical independence. Sky emission as observed by the WMAP satellite contains essentially emission from CMB anisotropies, from the galactic interstellar medium (ISM), and from a small number of strong radiosources. Other emissions, such as the Sunyaev-Zel'dovich (SZ) emission from galaxy clusters, or the emission from a backgound of numerous and unresolved radio and infrared sources, are too faint to be clearly seen in WMAP data.
In these proceedings, we discuss briefly some recent work on the extraction of a clean CMB map, of maps of foreground emission, and of the detection of thermal SZ effect towards a set of 893 known ROSAT clusters.

\section{A clean CMB map from WMAP data}

The CMB is a very special component. Its spectral emission law is perfectly known theoretically. Its anisotropies are known to be very close to Gaussian, and are not significantly correlated to other emissions (with the exception of a small correlation with tracers of large scale structure such as galaxies and clusters of galaxies, as such large scale structure can induce CMB anisotropies via the Integrated Sachs-Wolfe effect).

The simplest model of multifrequency CMB observations such as those of the WMAP satellite is $\vx = \va s +  \vn$,
where $\vx$ is the vector of observations (e.g. five maps), $\va$ is the response to the CMB for all observations (e.g. a vector with 5 entries equal to 1 if WMAP data only are considered) and $\vn$ is the noise (including any foreground contaminants). Here it is assumed (for the moment) that all observations are at the same resolution.
Assuming that we know the noise covariance $\tR_n$, the (generalised) least square (GLS) solution gives an estimate $\hat{s}_{\rm GLS}$ of $s$ as follows:
\begin{equation}
\hat{s}_{\rm GLS} = \frac{\va^t \, {{\tR}_n}^{-1}}{\va^t \, {{\tR}_n}^{-1} \, \va} \, \vx = \frac{\va^t \, {{\tR}_x}^{-1}}{\va^t \, {{\tR}_x}^{-1} \, \va} \, \vx
\label{eq:GLS}
\end{equation}
where the noise covariance matrix $\tR_n$ is a $5 \times 5$ matrix, and $\tR_x$ is the covariance matrix of the total signal including CMB. The last equality comes from the fact that, recalling that $\tR_x = \tR_n + \sigma^2 \va \va^t$, where $\sigma^2$ is the variance of the CMB, and making use of the inversion formula:
\begin{equation}
  \left [ \tR_n + \sigma^2 \va \va^t \right ] ^{-1}
  = \tR_n^{-1} - \sigma^2 \frac{\tR_n^{-1} \va \va^t \tR_n^{-1}}{1+ \sigma^2 \va^t \tR_n^{-1} \va}
\end{equation}
we have $\va^t \, {{\tR}_x}^{-1} \propto \va^t \, {{\tR}_n}^{-1}$. 

In the absence of prior information about the foregrounds, the GLS solution cannot be implemented using ${{\tR}_n}^{-1}$ in equation \ref{eq:GLS}. Diagonal terms in $\tR_n$ contain contributions from instrumental noise and from foregrounds, whereas off-diagonal terms arise from the covariance of the foreground emission between WMAP channels, which is not known {\emph{a priori}}.

An alternate solution is to use the so-called Internal Linear Combination (ILC) method, which estimates the CMB as the linear combination of the input maps with minimum variance (in fact, minimum power) and unit response to CMB. The ILC CMB estimate is:
\begin{equation}
\hat{s}_{\rm ILC} = \frac{\va^t \, {\widehat{\tR}_x}^{-1}}{\va^t \, {\widehat{\tR}_x}^{-1} \, \va} \, \vx
\label{eq:ILC}
\end{equation}
where $\widehat{\tR}_x$ is the empirical covariance matrix of the observations, i.e. an estimate of the true covariance ${\tR}_x$, estimated on the observed maps.
Hence, the ILC method is, in some sense, an approximation of the GLS method. The ILC uses, instead of the unknown $\tR_n$ or $\tR_x$, the empirical covariance matrix $\widehat{\tR}_x$ of the observations.

\begin{figure}
\begin{center}
\psfig{figure=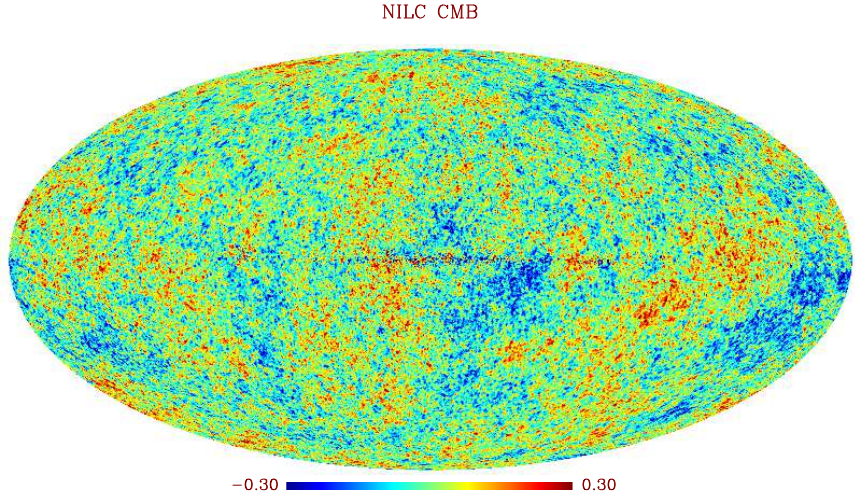,height=4.5in,angle=90}
\caption{CMB map obtained with WMAP 5 year data by needlet ILC (units are mK thermodynamic).
\label{fig:cmb}}
\end{center}
\end{figure}

An ILC estimate of the CMB map from WMAP data has been obtained by a number of authors. The ILC solution is just a linear combination of the inputs. However, it is clear that the optimal linear combination should depend on scale as well as on the region of the sky considered. This can be achieved by implementing the ILC using wavelets of some sort on the sphere. Here we use a decomposition on a needlet frame. The initial observations (WMAP  and a 100 micron map tracing the dust emission) are decomposed in sets of needlet coefficients, for a set of scales and a set of pixels on the sphere. The 100 micron observation helps subtracting dust emission on small scales, which is otherwise essentially significant in the W channel only (and hence cannot be removed by linear combinations of WMAP channels alone). The ILC is then implemented independently on sets of such coefficients, and a final CMB map, at full WMAP W channel resolution, is obtained by reconstructing the full sky CMB from the needlets coefficients of the CMB map obtained in this way \cite{2009A&A...493..835D}.
The resulting map is displayed in figure \ref{fig:cmb}. Comparison with earlier work shows that the needlet ILC (NILC) map is less contaminated by galactic foregrounds than most other ILC maps obtained by various authors without making use of the needlet decomposition. The NILC CMB map can be downloaded from a dedicated web page \footnote{{\sf http://www.apc.univ-paris7.fr/APC/Recherche/Adamis/cmb\_wmap-en.php}}.

\section{Foreground emission in WMAP data}

\begin{figure}[tb]
\begin{center}
\begin{tabular}{cc}
\psfig{figure=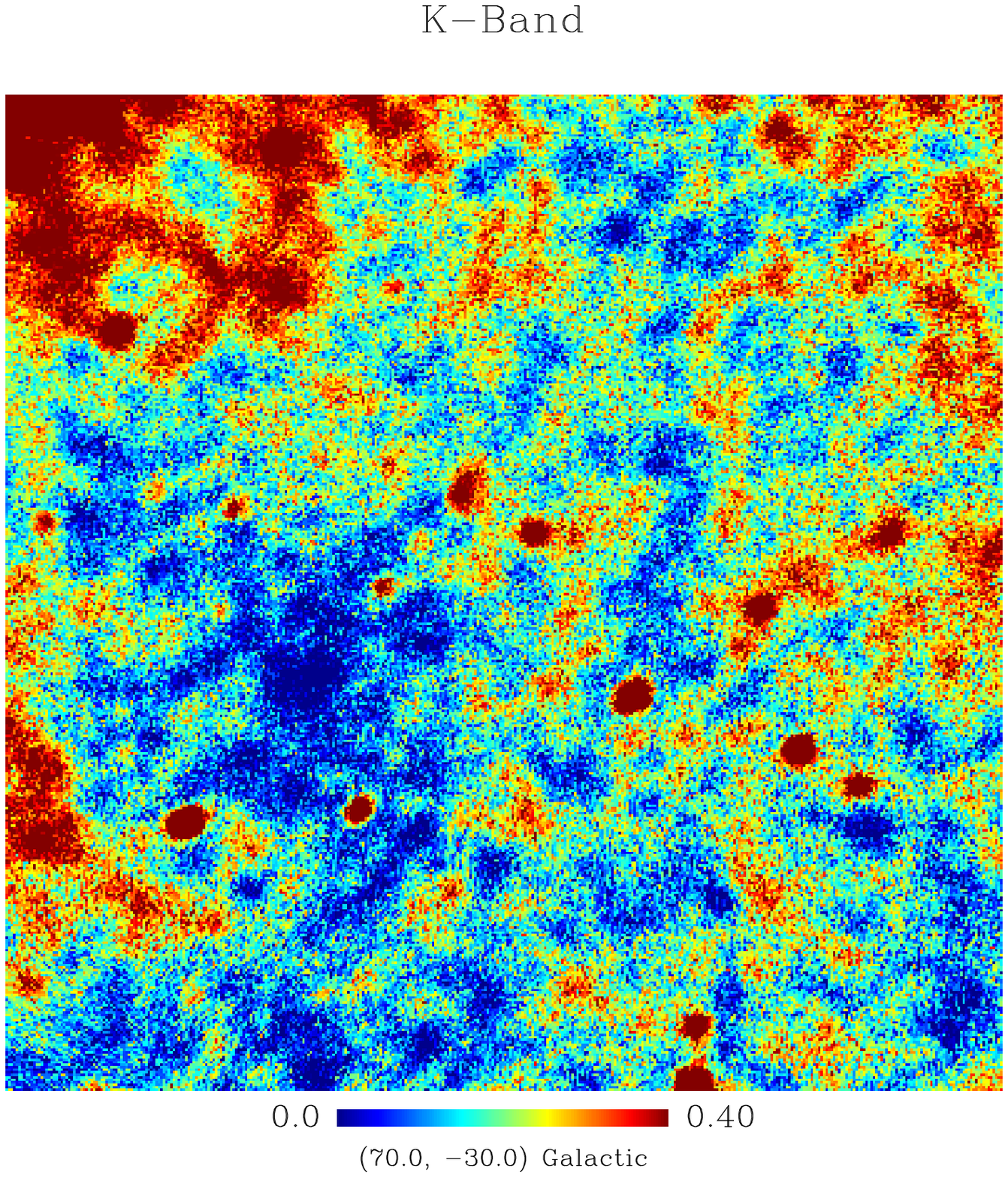,height=2.2in} & \psfig{figure=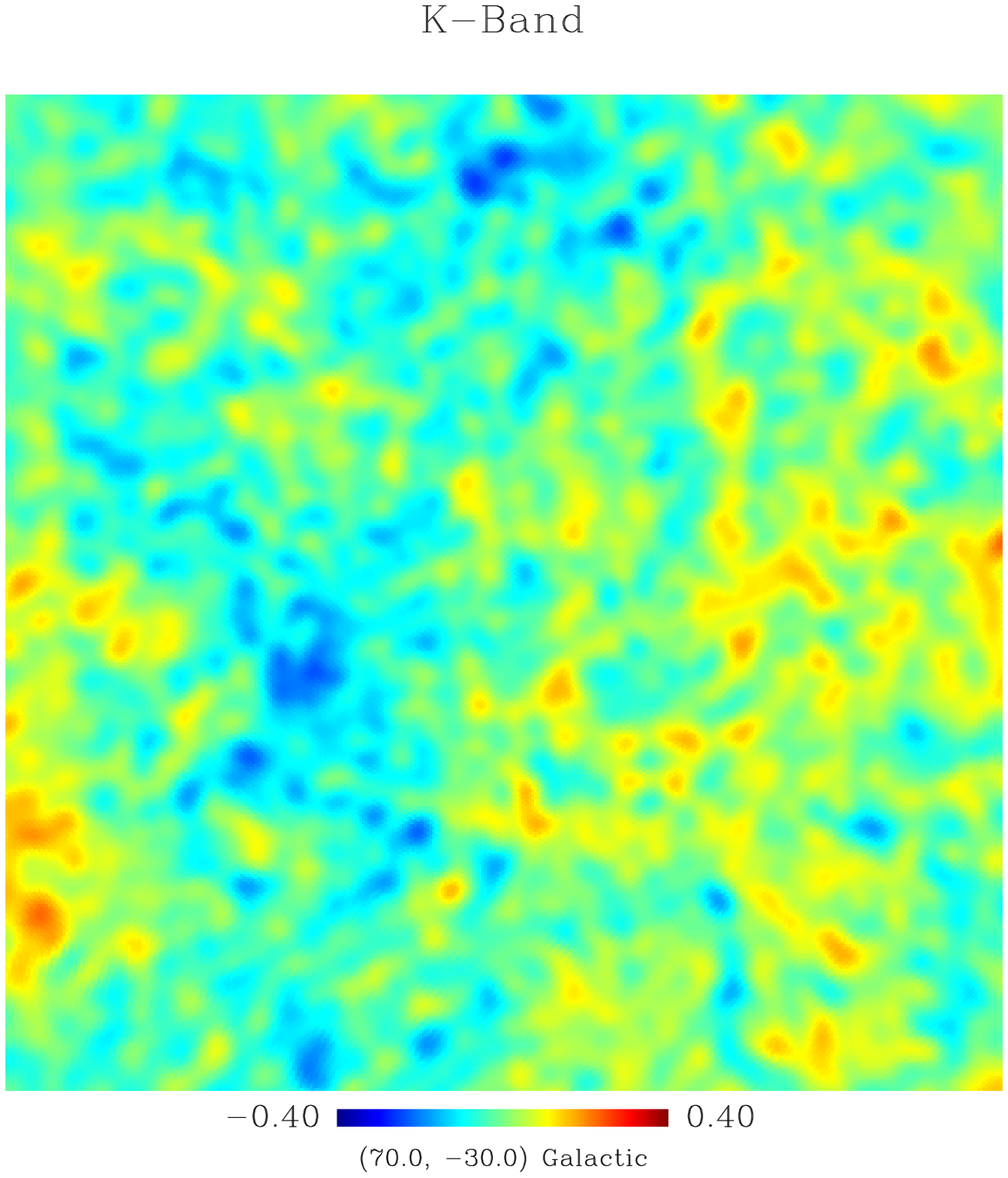,height=2.2in}\\
\psfig{figure=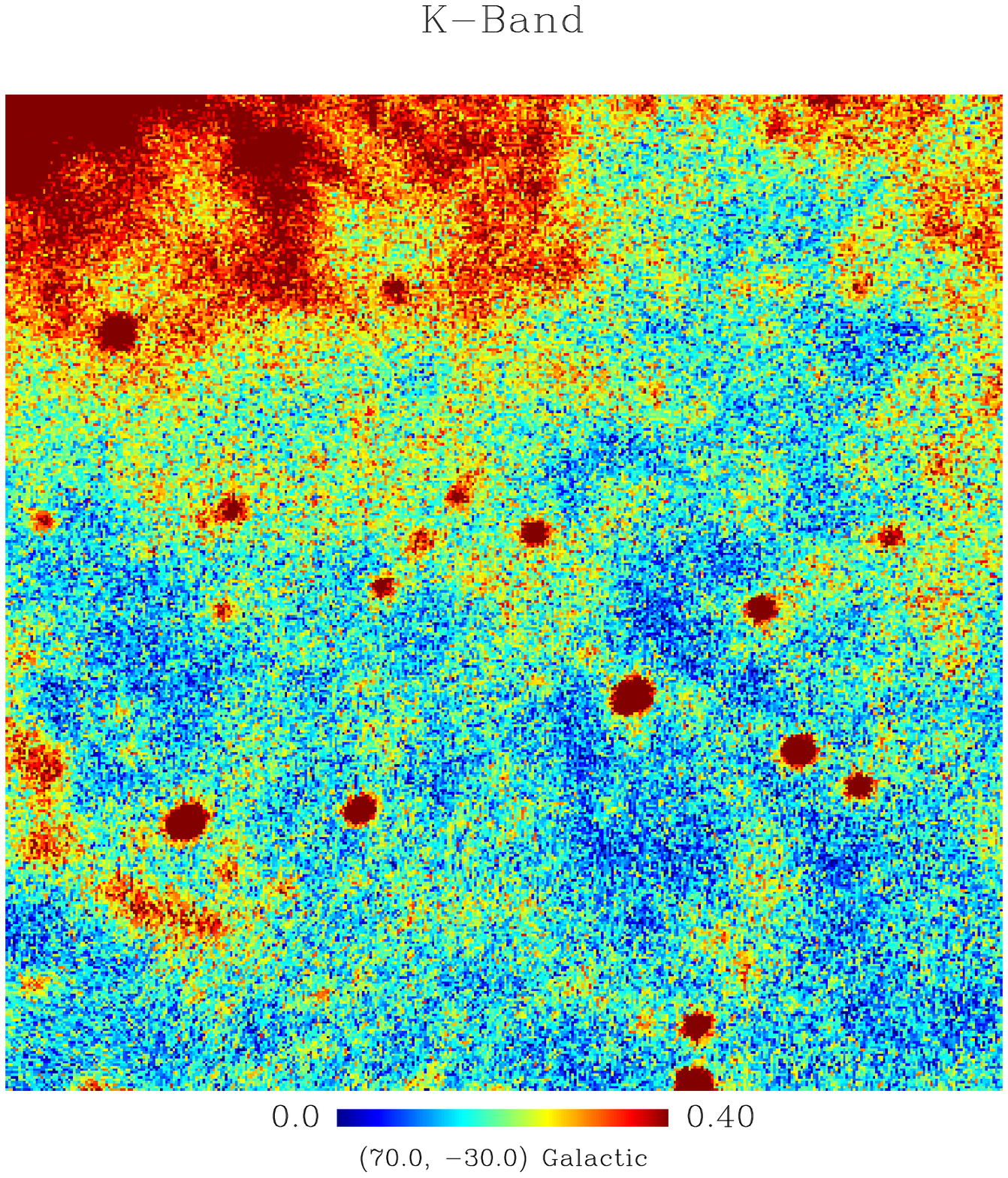,height=2.2in} & \psfig{figure=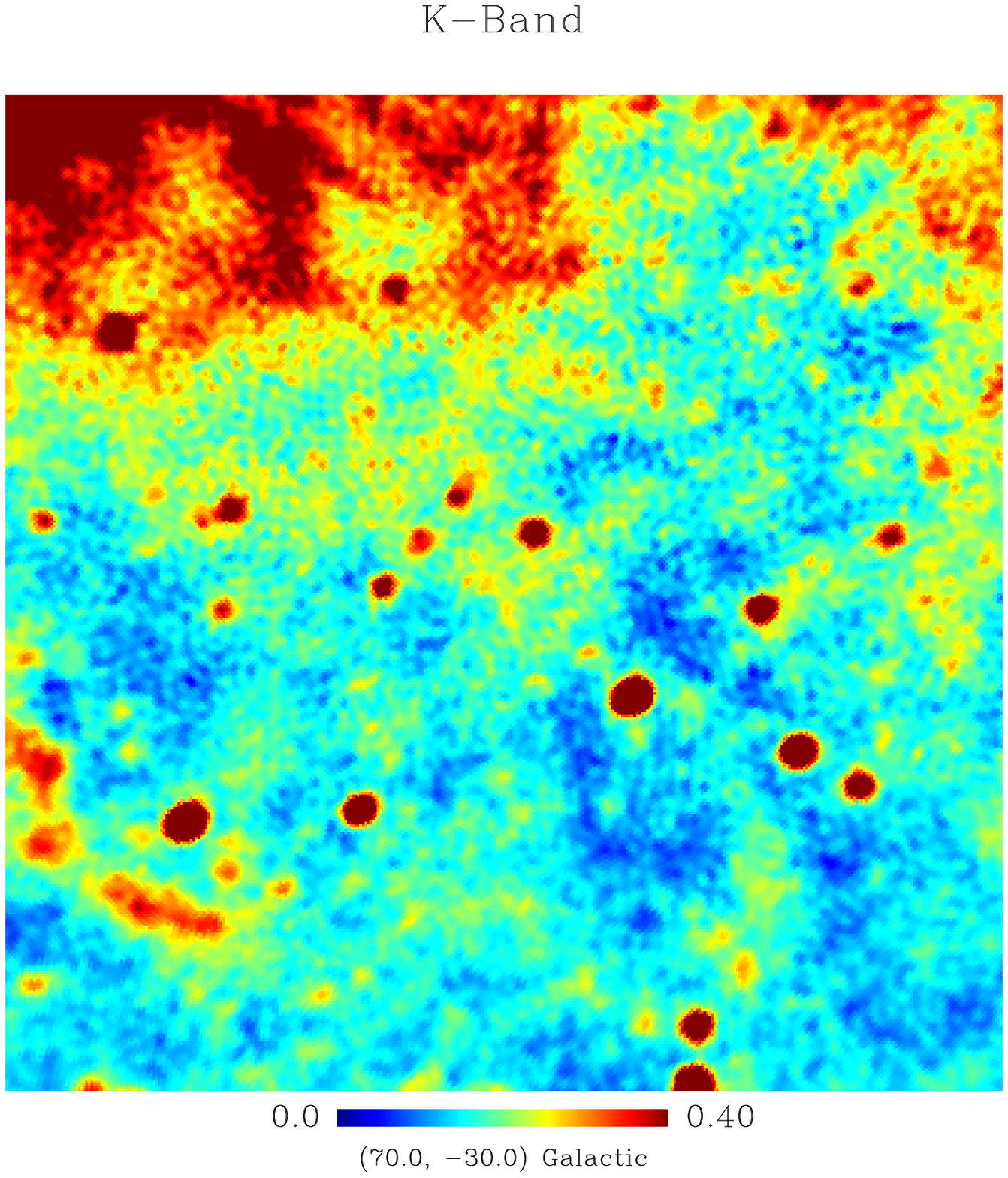,height=2.2in}\\
\end{tabular}
\caption{Top: left = K-band map, right = CMB. Bottom: left = after CMB subtraction, right = after filtering.}
\label{fig:demo}
\end{center}
\end{figure}

Once a clean CMB map has been obtained, it is quite natural to subtract it from the original observations to get foreground maps free of CMB contamination. It should be noted, however, that on small scales the CMB map at the resolution of the WMAP W channel contains more noise than CMB. Direct subtraction of the CMB map from the observation of one WMAP channel, with simple smoothing to put it at the resolution of the channel considered, would result in significant small scale extra noise. This noise, in addition, would be correlated from channel to channel. Hence, we subtract from each band-averaged map a Wiener-filtered version of the NILC CMB map \cite{ghosh_etal}.
After this is done, latitude-dependent additional smoothing, using a Gaussian beam which maximizes the signal to noise ratio at that latitude, is performed. The impact of the various stage of the processing for a patch of the WMAP K-band map at moderate galactic latitude ($l=70^\circ$, $b=-30^\circ$) is illustrated in figure \ref{fig:demo}.

\section{Cluster SZ emission in WMAP data}

The thermal SZ effect, due to inverse Compton scattering of CMB electrons off hot electrons in intra-cluster gas, leaves in principle a detectable (albeit small) inprint in CMB observations. WMAP, however, is not an ideal instrument for its detection. The pixel sensitivity is not quite good enough, and the angular resolution of 15 to 60 arcminutes (depending on the channel) does not permit to resolve the large majority of galaxy clusters.

Given a cluster profile and the WMAP beam profiles for all frequency bands, one can build multifrequency matched filters \cite{melin_mmf} to detect clusters in WMAP data. A blind search does not yield any significant detection. It is possible, however, to look for known SZ clusters observed by ROSAT, using a prior model profile derived from X-ray observations for each of them. The detection does  not depend much on this profile, as for unresolved clusters the actual cluster profile on the maps is essentially set by the beam.

For each individual ROSAT cluster, no significant SZ signal is detected, even when making use of prior information about the cluster location and profile. It is possible, however, to average the estimated SZ flux (the $Y_{\rm SZ}$ parameter) from clusters in redshift bins, in bins of X-ray luminosity, or in bins of mass (the mass being estimated from the X-ray flux). The SZ signal is then clearly detected. This permits to measure how the SZ flux depends on the cluster mass and on the X-ray luminosity. Hence, even if clusters are not detected individually, a filter which makes use both of the frequency dependence and of the expected profile of the clusters permits to extract relevant information about a component which is, in this data set, very subdominant. The reader is invited to look for additional details in the relevant publication \cite{melin_etal}.

\section{Conclusion}
Component separation is an important aspect of the analysis of CMB observations. Examples of extraction of the CMB, foreground emission, and thermal SZ effect from WMAP data have been discussed. No single component separation method is optimal for the extraction of all components. Instead, one has to design methods adapted to the component(s) of interest and to the scientific objectives. A component separation pipeline must chain several such methods and put them all in a coherent frame. This is becoming crucial for upcoming sensitive experiments, for which instrumental noise will be a subdominant source of error as compared to the confusion due to emissions from a large number of astrophysical processes.

\end{document}